\documentclass[a4paper,12pt]{article}
\pdfoutput=1

\usepackage{epsfig}
\usepackage{amsmath}
\usepackage{amssymb}
\usepackage{bbm}
\usepackage{verbatim}
\usepackage{bm}
\usepackage{color}
\usepackage{dsfont}
\usepackage{caption}
\usepackage[subrefformat=parens,labelformat=parens]{subfig}
\usepackage{subfloat}
\usepackage{cancel}
\usepackage{cite}
\numberwithin{equation}{section}
\usepackage{booktabs}
\usepackage[utf8]{inputenc}
\usepackage[linktoc=all]{hyperref}

\def\re{\mbox{Re}\,}

\def\be{\begin{equation}}
\def\ee{\end{equation}}
\def\bea{\begin{eqnarray}}
\def\eea{\end{eqnarray}}
\def\bes{\begin{subequations}}
\def\ees{\end{subequations}}

\begin{document}

\rightline{ IFT-UAM/CSIC-16-062}

\vskip 1cm

\begin{center}
{\LARGE\bf Challenges for D-brane large-field inflation \\[3mm] with stabilizer fields} 

\vskip 1.5cm

{\large Aitor Landete,$^{a}$ Fernando Marchesano$^{a}$ and Clemens Wieck$^{a,b}$}\\[10mm]
{\it{
${}^a$ Instituto de F\'{\i}sica Te\'orica UAM-CSIC, Cantoblanco, 28049 Madrid, Spain \\[2mm] 
${}^b$ Departamento de F\'{\i}sica Te\'orica, 
Universidad Aut\'onoma de Madrid, 
28049 Madrid, Spain

}}
\end{center}

\vskip 1cm

\begin{abstract}
\noindent 
We study possible string theory compactifications which, in the low-energy limit, describe chaotic inflation with a stabilizer field. We first analyze type IIA setups where the inflationary potential arises from a D6-brane wrapping an internal three-cycle, and where the stabilizer field is either an open-string or bulk K\"ahler modulus. We find that after integrating out the relevant closed-string moduli consistently, tachyonic directions arise during inflation which cannot be lifted. This is ultimately due to the shift symmetries of the type IIA K\"ahler potential at large compactification volume. This motivates us to search for stabilizer candidates in the complex structure sector of type IIB orientifolds, since these fields couple to D7-brane Wilson lines and their shift symmetries are generically broken away from the large complex structure limit. However, we find that in these setups the challenge is to obtain the necessary hierarchy between the inflationary and Kaluza-Klein scales.

\end{abstract}

\thispagestyle{empty}

\newpage

\tableofcontents

%
%%
%%%
%%%%
%%%%%%%%%%%%%%%%%%%%%%%%%%%%%%%%%%%%%%%%%%%%%%%
%%%%%%%%%%%%%%%%%%%%%%%%%%%%%%%%%%%%%%%%%%%%%%%
%%%%%%%%%%%%%%				      %%%%%%%%%%%%%%%%%%%%%%
%%%%%%%%%%%%%%         Introduction         %%%%%%%%%%%%%%%%%%%%%%
%%%%%%%%%%%%%%				      %%%%%%%%%%%%%%%%%%%%%%
%%%%%%%%%%%%%%%%%%%%%%%%%%%%%%%%%%%%%%%%%%%%%%%
%%%%%%%%%%%%%%%%%%%%%%%%%%%%%%%%%%%%%%%%%%%%%%%

\section{Introduction}

One of the current pressing problems in string cosmology is how to build successful models of large-field inflation \cite{Baumann:2014nda,Westphal:2014ana,Silverstein:2016ggb}. A typical challenge is to decouple the inflaton sector from the rest of the fields that obtain a mass below the string and Kaluza-Klein scales. On the one hand, one needs an inflaton candidate which is several orders of magnitude lighter than these massive fields. On the other hand, one needs to make sure that the interaction with the heavier fields does not spoil the ability of the scalar potential to generate 50 or 60 $e$-folds of inflation. 

Circumventing this challenge is arguably one of the main issues for those string models that implement the idea of axion monodromy \cite{Silverstein:2008sg}. Early proposals addressed the problem of creating a mass hierarchy by considering a source for the inflaton potential different from other sources of moduli stabilization, namely a brane-anti-brane system, and then using warping effects to lower the inflaton mass \cite{McAllister:2008hb,Berg:2009tg,Palti:2014kza,Retolaza:2015sta}. Recently it has been realized that a more systematic treatment can be applied in the context of F-term axion monodromy \cite{Marchesano:2014mla,Blumenhagen:2014gta,Hebecker:2014eua}. Indeed, in this framework both the inflaton and moduli stabilization potentials are described by a single 4d F-term supergravity potential, at least for small values of the inflaton field. Hence, one can apply the standard 4d supergravity techniques to describe the interplay of the inflaton sector with the rest of the heavy fields of the compactification. Even if this analysis is only valid at small inflaton values, it provides a first non-trivial test for the viability of the different F-term axion monodromy models proposed in the literature \cite{Marchesano:2014mla,Blumenhagen:2014gta,Hebecker:2014eua,Ibanez:2014kia,Hassler:2014mla,McAllister:2014mpa,Franco:2014hsa,Dudas:2014pva,Blumenhagen:2014nba,Hayashi:2014aua,Hebecker:2014kva,Ibanez:2014swa,Garcia-Etxebarria:2014wla,Hebecker:2015rya,Escobar:2015fda,Blumenhagen:2015xpa,Escobar:2015ckf,Baume:2016psm}.

In cases where the inflaton candidate is a complex structure modulus in type IIB flux compactifications, such techniques were applied in \cite{Blumenhagen:2014nba,Hayashi:2014aua,Hebecker:2014kva}. It was found that the required mass hierarchy between the inflaton and heavier moduli is in general hard to obtain, and that since the source of the inflaton potential and moduli stabilization is the same---that is, background fluxes---both systems are hard to decouple. Based on this \cite{Escobar:2015fda} proposed to generate the F-term inflaton potential by a different source, namely the presence of a D-brane. Similar to \cite{McAllister:2008hb,Berg:2009tg,Palti:2014kza,Retolaza:2015sta} this could allow to separate the inflaton mass scale from the remaining moduli by means of warping effects, as discussed in \cite{Escobar:2015ckf} for the case of type IIA flux compactifications with D6-branes. Moreover, the inflaton sector of this scenario is very similar to the 4d supergravity model of chaotic inflation proposed in \cite{Kawasaki:2000yn}, featuring a bilinear superpotential of the form
\be
W_\text{inf}\, =\, \Phi S\,,
\label{Winfintro}
\ee
where $\Phi$ contains the inflaton and $S$ is the so-called stabilizer field, the expectation value of the latter vanishing during the whole inflationary process.\footnote{See \cite{Dudas:2014pva,Hayashi:2014aua} for previous attempts to realize such a superpotential in type II compactifications.} Microscopically, one of the fields in this superpotential is a closed-string mode and the other one is an open-string mode \cite{Marchesano:2014iea}. As a result, the mass hierarchy between the inflaton and the moduli sectors is achieved by means of a hierarchy between closed- and open-string kinetic terms \cite{Escobar:2015ckf}, warping effects being particularly important for the latter. 

In the present paper we elaborate on this scenario further by considering the backreaction of heavy fields on the inflationary sector, following the prescription of \cite{Buchmuller:2014vda}. We find that the whole system is very sensitive to the presence of continuous shift symmetries in the K\"ahler potential. Indeed, on the one hand, the K\"ahler potential must be shift-symmetric in one of the components of $\Phi$ for this component to be a viable inflaton candidate. On the other hand, no shift symmetry should be present for $S$ or this field develops a tachyonic direction when taking into account the backreaction of heavy moduli during inflation. Due to this tachyonic direction $S$ attains an expectation value and the system fails to produce the desired period of slow-roll inflation. 

While this simple observation is not very constraining from the perspective of 4d supergravity model building \cite{Kallosh:2010xz}, it turns out to have important consequences for embedding the stabilizer scenario into string theory in the way described above. Indeed, in the case of type IIA string models all moduli stabilization scenarios have been formulated in the regime of large compactification volumes \cite{DeWolfe:2005uu,Camara:2005dc,Palti:2008mg}. As recently discussed in \cite{Carta:2016ynn}---see also \cite{Grimm:2011dx,Kerstan:2011dy}---in that regime the tree-level K\"ahler potential exhibits shift symmetry in both closed- and open-string modes. As a result, we find that in the D6-brane scenario of \cite{Escobar:2015ckf} none of the chiral fields involved in (\ref{Winfintro}) can play the role of a stabilizer field. 

Of course, continuous shift symmetries are not ubiquitous in string theory and are in fact broken by different effects. One may then consider a scenario where the superpotential (\ref{Winfintro}) is still realized but no shift symmetries appear for one of the two fields. In the context of moduli stabilization, the obvious choice would be to identify $S$ with a complex structure modulus away from the large complex structure limit, where shift symmetries are generically broken. However, following \cite{Marchesano:2014iea}, we find that then $\Phi$ must be given by a D7-brane Wilson line and, by the results of \cite{Marchesano:2008rg}, that the kinetic terms of these modes are unaffected by warping. Therefore, contrary to the previous type IIA scenario, here we find no obvious obstruction to stabilizing the inflationary trajectory, but also no mechanism to decouple the inflaton sector from the moduli and Kaluza-Klein scales. 

The rest of this paper is organized as follows. In Section \ref{sec:gen} we discuss how supersymmetrically integrating out heavy moduli is, to leading order, equivalent to replacing them by their vacuum expectation values in the K\"ahler and superpotential. In Section~\ref{sec:D6back} we apply this observation to type IIA D6-brane models of large-field inflation. We find that the shift symmetries present in the tree-level K\"ahler potential destabilize the inflationary trajectory. In Section \ref{sec:new} we turn to analyze a similar system in the context of type IIB compactifications without shift symmetries in the complex structure sector. However, here we find no obvious way to hierarchically lower the inflaton mass with respect to the Kaluza-Klein scale. We summarize our results and draw our conclusions in Section \ref{sec:syc}.

%
%%
%%%
%%%%
%%%%%%%%%%%%%%%%%%%%%%%%%%%%%%%%%%%%%%%%%%%%%%%
%%%%%%%%%%%%%%%%%%%%%%%%%%%%%%%%%%%%%%%%%%%%%%%
%%%%%%%%%%%%%%				    %%%%%%%%%%%%%%%%%%%%%%
%%%%%%%%%%%%%%         Section 2           %%%%%%%%%%%%%%%%%%%%%%
%%%%%%%%%%%%%%				    %%%%%%%%%%%%%%%%%%%%%%
%%%%%%%%%%%%%%%%%%%%%%%%%%%%%%%%%%%%%%%%%%%%%%%
%%%%%%%%%%%%%%%%%%%%%%%%%%%%%%%%%%%%%%%%%%%%%%%

\section{Integrating out moduli supersymmetrically: A shortcut}
\label{sec:gen}

Describing inflation with low-energy effective string actions can often be split into two problems. On the one hand, obtaining a comparably light scalar field with a suitable scalar potential. The latter must be able to generate at least 50 to 60 $e$-folds of inflation at a characteristic scale $H$ in accordance with CMB measurements. On the other hand, stabilizing all remaining moduli in a Minkowski or de Sitter vacuum at a mass scale greater than $H$. Here we focus on the latter problem and its implications for the former. Moreover, we consider setups in which all moduli are stabilized in a supersymmetric Minkowski vacuum because large-field inflation with a stabilizer field actually forces us to do so, as discussed in \cite{Buchmuller:2014pla}. For our purposes it suffices to leave the precise mechanism of moduli stabilization unspecified, and instead assume the existence of a superpotential piece $W_\text{mod}(\rho_i) \subset W$ which satisfies $\langle D_{\rho_i} W_\text{mod} \rangle = 0$ for all relevant moduli fields $\rho_i$. Examples are known in the literature, they include the racetrack setup of \cite{Kallosh:2004yh} and a less fine-tuned mechanism using an additional stabilizer field \cite{Wieck:2014xxa}.\footnote{Note that supersymmetry is necessarily broken in the original setup of \cite{Kachru:2003aw} once the vacuum is uplifted to a Minkowski or de Sitter background. The same applies to the extensions of \cite{Balasubramanian:2005zx} and \cite{Balasubramanian:2004uy,Westphal:2006tn}, in which the breaking scale is typically very high.}

In many string-effective inflation models the inflaton and the moduli interact even if the moduli are much heavier than the dynamical scale of inflation. Through supergravity couplings this even happens in models where the superpotential separates, 
\begin{align}
{W = W_\text{inf} (\phi_i) + W_\text{mod}(\rho_i)}\,,
\end{align}
where $\phi_i$ collectively denotes the superfields involved in the inflationary part of the theory. This interaction in the Lagrangian introduces what we call a ``backreaction'' of the heavy fields on the inflationary potentials. Many models of this type have been constructed in the recent literature, from various different corners of string theory. The effect of stabilizing and integrating out the fields $\rho_i$, i.e., the backreaction, has been systematically studied in \cite{Buchmuller:2014vda}. In cases where all $\rho_i$ appear logarithmically in the K\"ahler potential, the effective potential for the fields $\phi_i$ at leading order reduces to the scalar potential of the inflationary sector alone, as if the moduli had not been present as dynamical degrees of freedom. This is true as long as all moduli masses, determined by the second derivatives of $W_\text{mod} (\rho_i)$, lie above the Hubble scale $H$, determined by $W_\text{inf}$ and its first derivatives. Since $\langle D_{\rho_i} W_\text{mod} \rangle = 0$ this confirms a naive expectation fuelled by old QFT arguments: if they are heavy enough and do not break supersymmetry, the moduli completely decouple. This statement is true up to sub-leading corrections which arise in powers of $H/m_{\rho_i}$, cf.~\cite{Buchmuller:2014vda} for details. These corrections are under control whenever the moduli can be safely integrated out. Still they may be sizeable and lead to slightly changed predictions of a given model, such as the CMB observables. In particular, the higher-order terms arising in powers of $H/m_{\rho_i}$ lead to a flattening of the potential \cite{Dong:2010in,Buchmuller:2014vda,Buchmuller:2015oma}.

Despite the interesting effects that these corrections may have, in this paper we aim to analyze the stability of the inflationary trajectories after moduli backreaction, for which it suffices to focus on the leading-order result for the effective action. In \cite{Buchmuller:2014vda} and subsequent publications this has been obtained by computing the supergravity potential and solving the inflaton-dependent equations of motion for the moduli fields. Depending on the details of the setup, this can be a tedious exercise. Therefore we wish to point out here that the leading-order effective potential, taking moduli backreaction into account, can be obtained via a simple shortcut. The key observation is that integrating out the heavy $\rho_i$ is equivalent to fixing all $\rho_i$ in $W$ and $K$ at their expectation values in the vacuum, and subsequently computing the scalar potential for the remaining fields $\phi_i$. The result corresponds to the full effective potential in the limit $m_{\rho_i} \to \infty$. Clearly, however, corrections due to the finiteness of $m_{\rho_i}$---such as the flattening corrections mentioned above---cannot be obtained in this way.

%
%%
%%%
%%%%
%%%%%%%%%%%%%%%%%%%%%%%%%%%%%%%%%%%%%%%%%%%%%%%
%%%%%%%%%%%%%%%%%%%%%%%%%%%%%%%%%%%%%%%%%%%%%%%
\subsection{A no-scale toy model}
\label{sec:toy1}

Let us demonstrate this claim in a few simple examples. Consider a simple no-scale model with a single K\"ahler modulus $T$ and an inflaton multiplet $\Phi$,
\begin{align}\label{eq:Toy1}
K = - 3 \log \left[T + \bar T - \frac12 (\Phi + \bar \Phi)^2 \right] \,, \qquad W = \frac12 m \Phi^2 + W_\text{mod}(T)\,.
\end{align}
For a similar illustration this toy model has already been considered in \cite{Buchmuller:2014pla}. It corresponds to a boiled-down variant of some of the F-term axion monodromy models in the recent literature.\footnote{For the purposes of this discussion the precise form of $K$ does not matter. In particular, our results remain unchanged whether or not there is kinetic mixing between $\Phi$ and $T$. Moreover, for simplicity we assume all constant parameters in the superpotential to be real in this example.} The corresponding scalar potential reads
\begin{align}\label{eq:ToyV1}
V(\varphi, t) = \frac{1}{6 t} \left[ \left(\frac16 m^2 + \frac12 m W_\text{mod}^\prime (t) \right) \varphi^2 - \frac{3 W_\text{mod}(t) W_\text{mod}^\prime(t)}{t} +  W_\text{mod}^\prime(t)^2 \right]\,,
\end{align}
where $\varphi$ is the canonically normalized inflaton field and $t = \text{Re}\, T$. The other two real scalars do not play a role in this case and have been set to zero. They do not have linear terms in $V$ and do not displace the inflaton. Moreover, their masses are positive and large compared to $H$.

At first sight, this theory has a supersymmetric Minkowski vacuum at $t = t_0$ with $W_\text{mod}^\prime(t_0)=0$ and $W_\text{mod}(t_0)=0$. On the inflationary trajectory, then, \eqref{eq:ToyV1} reduces to a simple quadratic potential for $\varphi$. However, this is not really true because \eqref{eq:ToyV1} contains non-trivial interaction terms between $t$ and $\varphi$. In particular, minimizing the full potential with respect to $t$ leads to
\begin{align}
t_\text{min} \simeq t_0 - \frac{m \varphi^2}{4 W_\text{mod}^{\prime \prime}(t_0)} + \mathcal O\left( \frac{m^2 \varphi^2}{{W_\text{mod}^{\prime \prime}(t_0)}^2}\right)\,,
\end{align}
at leading order in powers of $H/m_t$, where $m_t \sim W_\text{mod}^{\prime \prime}(t_0) $. Plugging this back into \eqref{eq:ToyV1} leads to the proper effective potential for the inflaton,
\begin{align}
V(\varphi) = \frac{1}{18 t_0} \left( \frac12 m^2 \varphi^2 - \frac{3}{16} m^2 \varphi^4 \right) + \mathcal O\left( \frac{m \varphi}{W_\text{mod}^{\prime \prime}(t_0)}\right)\,.
\end{align}
Evidently, the interaction during inflation interferes with the cancellation of the negative definite term in the supergravity potential. Taking the backreaction of $t$ into account reintroduces the term proportional to $-3 |W|^2$, which makes the model fail. 

Most importantly, we could have seen this much faster. Instead of setting $t= t_0$ in the scalar potential, which leads to the wrong result, we must replace $T = t_0$ in $K$ and $W$ defined in \eqref{eq:Toy1}. Treating only $\Phi$ as dynamical, we observe that 
\begin{align}\label{eq:Toyeff}
K =  - 3 \log \left[2 t_0- \frac12 (\Phi + \bar \Phi)^2 \right] \,, \qquad W = \frac12 m \Phi^2 + W_\text{mod}(t_0)\,,
\end{align}
leads to the correct leading-order potential
\begin{align}
V(\varphi) = \frac{1}{18 t_0} \left( \frac12 m^2 \varphi^2 - \frac{3}{16} m^2 \varphi^4 \right)\,.
\end{align}
As stressed before, this simplified treatment corresponds to taking ${m_t \sim W_\text{mod}^{\prime \prime}(t_0) \to \infty}$, and thus it is insufficient for computing corrections.

%
%%
%%%
%%%%
%%%%%%%%%%%%%%%%%%%%%%%%%%%%%%%%%%%%%%%%%%%%%%%
%%%%%%%%%%%%%%%%%%%%%%%%%%%%%%%%%%%%%%%%%%%%%%%
\subsection{A no-scale toy model with stabilizer field}
\label{sec:toy2}

Let us even spend time on a second example which contains a stabilizer field $S$. While the latter is supposed to eliminate the dangerous term proportional to $-3 |W|^2$, effects of the moduli backreaction are important and can be observed using our shortcut. Consider
\begin{align}\label{eq:Toy2}
K = - 3 \log \left[T + \bar T - \frac12 (\Phi + \bar \Phi)^2 \right] + \frac12 (S + \bar S)^2\,,  \qquad W = m S \Phi + W_\text{mod}(T)\,,
\end{align}
which is a simplified version of some of the effective theories that arise in D-brane inflation, as explained in Section 3. Neglecting the explicitly modulus-dependent terms proportional to $W_\text{mod}$ and its first derivative for now, we find the following scalar potential.
\begin{align}
V(S, \varphi, t) = \frac{1}{12 t^2} \left[ \frac12 m^2 \varphi^2 + \frac12 (m^2 + 3 m^2 \varphi^2) s_1^2 + \frac12 m^2 s_2^2 + \mathcal O(W_\text{mod}(t), W_\text{mod}^\prime(t)) \right]\,,
\end{align}
where we have expanded in the relevant fields up to quadratic order. Notice that we have written $S = (s_1 + i s_2)/\sqrt2$. At this level the picture seems to be the following: $\varphi$, $s_1$, $s_2$ have equal supersymmetric masses. In addition, $s_1$ receives a supersymmetry-breaking mass term through its K\"ahler potential coupling to the inflationary vacuum energy. While $s_2$ is not heavy enough to satisfy a single-field treatment of inflation for arbitrary initial conditions, the model appears consistent. This would remain true if we naively set $t = t_0$, which entails $W_\text{mod}(t_0) = W_\text{mod}^\prime(t_0) =0$.

The consistency no longer holds when we take the backreaction of $t$ into account by setting $T = t_0$ in eqs.~\eqref{eq:Toy2}. What we find for the leading-order effective potential of the $S$ - $\Phi$ system, using
\begin{align}\label{eq:Toy3}
K = - 3 \log \left[2 t_0- \frac12 (\Phi + \bar \Phi)^2 \right]  + \frac12 (S + \bar S)^2\,,  \qquad W = m S \Phi + W_\text{mod}(t_0)\,,
\end{align}
is instead
\begin{align}
V(S,\varphi) = \frac{1}{12 t_0^2} \left[ \frac12 m^2 \varphi^2 + \frac12 \left(m^2 + \frac34 m^2 \varphi^2\right) s_1^2 + \frac12 \left(m^2 - \frac34 m^2 \varphi^2\right) s_2^2  \right]\,.
\end{align}
One can check that the same result is found after consistently minimizing $T = T_\text{min}(S,\varphi)$ during inflation. Notice that $s_2$ is actually a tachyonic direction during inflation. While $s_1$ is saved from the same fate by its soft mass term proportional to $H^2$, the model never yields successful slow-roll inflation due to the tachyonic direction along $s_2$. This is ultimately due to the shift symmetry of the stabilizer field, and was only concealed by a would-be no-scale cancellation in the modulus sector. As we explore in the next section, this is exactly what causes the D6-brane inflation model of \cite{Escobar:2015fda,Escobar:2015ckf} to fail.\footnote{In \cite{Dudas:2015lga} this phenomenon of a destructive backreaction was shown to arise in many other inflation models, involving stabilizer fields or not.}

%
%%
%%%
%%%%
%%%%%%%%%%%%%%%%%%%%%%%%%%%%%%%%%%%%%%%%%%%%%%%
%%%%%%%%%%%%%%%%%%%%%%%%%%%%%%%%%%%%%%%%%%%%%%%
%%%%%%%%%%%%%%				    %%%%%%%%%%%%%%%%%%%%%%
%%%%%%%%%%%%%%         Section 3           %%%%%%%%%%%%%%%%%%%%%%
%%%%%%%%%%%%%%				    %%%%%%%%%%%%%%%%%%%%%%
%%%%%%%%%%%%%%%%%%%%%%%%%%%%%%%%%%%%%%%%%%%%%%%
%%%%%%%%%%%%%%%%%%%%%%%%%%%%%%%%%%%%%%%%%%%%%%%

\section{D6-brane inflation and backreaction of closed-string moduli}
\label{sec:D6back}

In the following we would like to apply the general remarks of Section \ref{sec:gen} to examine string theory models of large-field inflation. In particular, in this section we focus on the proposal made in \cite{Escobar:2015fda,Escobar:2015ckf} to embed models with stabilizer fields in type IIA compactifications with D6-branes. As we will see, taking into account the shift symmetries of the model and applying the above shortcut to integrate out heavy fields leads to tachyonic directions within the inflationary system which, as in the toy model above, spoil slow-roll inflation. 

%
%%
%%%
%%%%
%%%%%%%%%%%%%%%%%%%%%%%%%%%%%%%%%%%%%%%%%%%%%%%
%%%%%%%%%%%%%%%%%%%%%%%%%%%%%%%%%%%%%%%%%%%%%%%
\subsection{D6-brane inflation}
\label{sec:D6}

In \cite{Escobar:2015fda} a new proposal to embed models of large-field inflation in string theory was made, based on the property of certain D-branes to generate bilinear superpotentials for open- and closed-string axions \cite{Marchesano:2014iea}. This was then developed in the context of type IIA string compactifications with D6-branes, as recently discussed in further detail in \cite{Escobar:2015ckf}. In essence the setup features a D6-brane that creates an inflationary potential for a B-field axion and the Wilson line of the brane. Near the supersymmetric vacuum the low-energy supergravity is that of chaotic inflation with a stabilizer field, as first proposed in \cite{Kawasaki:2000yn} and generalized in \cite{Kallosh:2010xz}. As discussed in \cite{Marchesano:2014iea} the D6-brane couples to the background in such a way that the following superpotential is developed
\begin{align}\label{eq:D6W}
W_\text{inf} = n_a T^a \Phi = T \Phi\,,
\end{align}
where $n_a \in \mathbb{Z}$, $\Phi$ is the superfield containing the D6-brane Wilson line, and $T = n_aT^a$ is a linear combination of K\"ahler moduli such that $b = \text{Im}\,T$ is the B-field axion that couples to the D6-brane. Following \cite{Kawasaki:2000yn} it is clear that such a superpotential can yield an effective description of chaotic inflation if at least one of the two chiral fields is light enough (usually through the appearance of a shift symmetry) and the other one is significantly heavier. 

The shift symmetries of this system can be analyzed through the effective K\"ahler potential for the closed- and open-string moduli in type IIA orientifold compactifications, first discussed in \cite{Grimm:2011dx,Kerstan:2011dy} and more recently in \cite{Carta:2016ynn}. There it was argued that $K = K_\text{K} + K_\text{Q}$, where on the one hand
\begin{equation}
\label{eq:D6KK}
K_\text{K} = - \log \left[ \frac{1}{6} \kappa_{abc}(T^a + \bar T^a)(T^b + \bar T^b)(T^c + \bar T^c) \right]\,,
\end{equation}
with $T^a$ the K\"ahler moduli of the compactification and $\kappa_{abc}$ the corresponding triple intersection numbers.\footnote{In order to connect with the standard notation in the 4d supergravity literature used in Section \ref{sec:gen}, our conventions differ from those in \cite{Grimm:2011dx,Kerstan:2011dy,Carta:2016ynn} and are such that $T^a = t^a + i b^a$, with $b^a$ the B-field axions of the compactification. The same applies to the complex structure moduli, with $\text{Im}\,U'^K$ containing the axionic piece of the field.} On the other hand, for a choice of Calabi-Yau three-form symplectic basis we can write $K_\text Q$ as \cite{Kerstan:2011dy} 
\be
K_\text Q = - 2 \log \left( \frac{1}{16} {\cal F}_{KL} \left( U'^K + \bar{U}'^K\right) \left( U'^L + \bar{U}'^L\right)\right)\,,
\label{eq:D6KQp}
\ee
where $\text{Re}\, U'^K$ are defined in terms of the periods of the three-form ${\rm Re}\, \Omega$, and  ${\cal F}_{KL}$ are real functions that only depend on their quotients, such that they are invariant under the overall rescaling $U'^K \rightarrow \lambda U'^K$. The most involved part in describing $K_\text Q$ is determining how the geometric quantities $U'^K$ depend on the holomorphic variables of the four-dimensional effective theory. By the analysis of \cite{Carta:2016ynn} one obtains that
\be
U'^K\, =\, U^K + \frac{1}{2} T^a {\bf H}_a^K\,,
\ee
where $U^K$ is the new holomorphic variable and ${\bf H}_a^K$ a homogeneous function of degree zero in $\text{Re}\, T^a$, $\text{Re}\, \Phi$, and $\text{Re}\, U^K$. The leading-order term is of the form 
\be
{\bf H}_a^K\, =\, - \frac{1}{2} Q^K \eta_a \frac{(\Phi + \bar{\Phi})^2}{[\eta_a (T^a + \bar{T}^a)]^2} + \dots \,,
\ee
where $Q^K$ and $\eta_a$ can be taken to be constants that depend on the D6-brane embedding. Putting all this together we obtain the following approximate expression,
\be
\label{eq:D6KQ}
K_\text{Q} = -2 \log\left\{ \frac{1}{16} \mathcal F_{KL} \left[ U^K + \bar U^K - \frac{1}{8} \tilde{Q}^K (\Phi + \bar \Phi)^2 \right] \left[ U^L + \bar U^L - \frac{1}{8} \tilde Q^L (\Phi + \bar \Phi)^2 \right] + \dots\right\}\
\ee
where we have defined $\tilde Q^K = Q^K/(\eta_a \text{Re}\, T^a)$. This expression for $K_\text Q$ resembles the one used in \cite{Escobar:2015fda,Escobar:2015ckf} except for the fact that here $\tilde Q^K$ is moduli-dependent. This is not  important when applying the philosophy of Section \ref{sec:gen}, since upon integrating out all closed-string moduli except $T$ we obtain an effective K\"ahler potential where the $\tilde Q^K$ become constants.\footnote{When $\tilde Q^K$ also depends on the stabilizer field $T$ the discussion is a bit more involved. The coupling of $\Phi$ and $T$ in $K$ introduces additional interactions in the scalar potential. However, one can check that these interaction terms arise first at $\mathcal O(T^3)$ in the action, which makes them irrelevant to the following discussion. We can thus safely treat $\tilde Q^K$ as constants in this case as well.} 

Notice that the K\"ahler potential only depends on $\text{Re} \, T^a$, $\text{Re} \, U^K$, and $\text{Re} \, \Phi$ and therefore it displays several shift symmetries. This is true in general, even without the simplifying assumptions that took us to the expression (\ref{eq:D6KQ}), and it only relies on considering type IIA at large compactification volumes compared to the string scale \cite{Carta:2016ynn}. These shift symmetries imply that, in principle, either $\text{Im} \, T$ or $\text{Im} \, \Phi$ could play the role of the inflaton field; both scenarios have been considered in \cite{Escobar:2015ckf}. Unfortunately this also means that the other field cannot play the role of the stabilizer field, a fact missed in the analysis of \cite{Escobar:2015ckf} where the backreaction of the heavy closed-string moduli was not taken into account. To see this point in detail we analyze the scalar potential for the inflaton system first from the viewpoint of \cite{Escobar:2015ckf}. Then, in Section \ref{sec:back}, we revisit the scalar potential by applying the philosophy of Section \ref{sec:gen} to see how backreaction destabilizes the inflationary trajectory.

%
%%
%%%
%%%%
%%%%%%%%%%%%%%%%%%%%%%%%%%%%%%%%%%%%%%%%%%%%%%%
\subsubsection*{The scalar potential without backreaction}

Let us consider the scenario in which the D6-brane Wilson line $\phi = \text{Im}\,\Phi$ is the inflaton candidate, and so $\text{Re}\,\Phi = T = 0$ defines the would-be inflationary trajectory. On this trajectory the superpotential \eqref{eq:D6W} generates a quadratic potential for $\phi$. The pressing issue at hand, however, is the stabilization of the closed-string moduli $U^K$ and $T^\alpha$, where the index $\alpha$ runs over all the K\"ahler moduli except $T$. In order to implement such a stabilization, $W_\text{inf}$ must be accompanied by an additional piece $W_\text{mod} (U^K, T^\alpha)$ which lifts the corresponding flat and run-away directions.\footnote{In general, $W_\text{mod}$ may also depend on $\Phi$ through the contribution of world-sheet and D2-brane instantons to the superpotential \cite{Escobar:2015ckf}. For our stability analysis we assume that such terms are negligible compared to the perturbative piece of $W$.}
 As in \cite{Escobar:2015fda,Escobar:2015ckf} we consider the case where none of these moduli break supersymmetry in the vacuum, that is when
\begin{align}\label{eq:Fvanish}
D_{U^K} W_\text{mod}\big|_{\Phi = 0} = D_{T^\alpha} W_\text{mod}\big|_{\Phi = 0} = 0\,,
\end{align}
and then expand the full F-term scalar potential around the inflationary trajectory to find an effective potential for $T$ and $\Phi$. In \cite{Buchmuller:2014pla} it was shown that \eqref{eq:Fvanish} is actually a necessary assumption in these kinds of setups: allowing the moduli to break supersymmetry in the vacuum leads to additional terms, essentially soft terms, proportional to $\langle W_\text{mod} \rangle$ and $\langle W^\prime_\text{mod} \rangle$. If one of them, or equivalently the scale of supersymmetry breaking, becomes too large the model fails due to a backreaction of the stabilizer field~$T$.

At quadratic order in the fields the resulting scalar potential of \cite{Escobar:2015ckf} reads\footnote{Here we exhibit the result obtained in \cite{Escobar:2015ckf}, which assumed a K\"ahler potential of the form \eqref{eq:D6KQ} and, following \cite{Kerstan:2011dy}, that $\tilde Q^K$ are moduli-independent. Had we taken into account the correct moduli dependence of these quantities and applied the same procedure a scalar potential different from \eqref{eq:V1} would have been obtained, although the subsequent discussion based on it would have been similar. The fact that the calculation of \cite{Escobar:2015ckf} yields different effective scalar potentials after changing the dependence of heavy fields in the initial K\"ahler potential indicates that the backreaction of the heavy fields cannot be neglected.}
\begin{align}\label{eq:V1}
V = e^K \left[ K^{\Phi \bar \Phi} |\partial_\Phi W_\text{inf}|^2 + K^{T \bar T}|\partial_T W_\text{inf} + T \partial_T^2 W_\text{mod}| + 4 (\text{Im}\,T)^2 (\text{Im}\,\Phi)^2 \right]\,,
\end{align}
where we have assumed that $W_\text{mod}$ is very small or vanishing at the vacuum. Taking the potential \eqref{eq:V1} at face value one can show that $\text{Re}\,\Phi$ and both components of $T$ have masses parametrically larger than the Hubble scale $H$, which means they can be safely integrated out during inflation, leading to the desired quadratic potential for $\phi$. Note that for $b = \text{Im}\,T$ this is due to the last piece in \eqref{eq:V1}, which appears as a remnant of the no-scale symmetry in the closed-string sector. In terms of canonically normalized fields \eqref{eq:V1} reads
\begin{align}\label{eq:V2}
V = \frac12 m^2 \varphi^2 + \left( \frac12 m^2 + m^2 \varphi^2\right) \sigma^2+ \left( \frac12 m^2 + 2 m^2 \varphi^2\right) t_1^2 + \left( \frac12 m^2 + \frac83 m^2 \varphi^2\right) t_2^2\,,
\end{align}
where $t_1$ and $t_2$ are the  components of the stabilizer field, $\varphi$ denotes the canonically normalized inflaton field, and $\sigma$ its saxionic partner. In this form the scalar potential mostly depends on the mass parameter $m$, which in turn depends on the constants in $K$, the volume, and the warping of the compact manifold. In this form the desired mass hierarchy $m_\varphi \ll m_\sigma, m_{t_1}, m_{t_2}$ during inflation is evident.

Finally, we may also consider the scenario where we take $b = \text{Im}\,T$ to be the inflaton candidate. Applying the approach of \cite{Escobar:2015ckf} and expanding the F-term potential along the new inflationary trajectory $\text{Re}\,T = \Phi  = 0$, we obtain a 
similar scalar potential but with the roles of $\Phi$ and $T$ exchanged. More precisely, we obtain \eqref{eq:V1} but with the interchange $\Phi \leftrightarrow T$. Needless to say, this leads to the same potential \eqref{eq:V2} for canonically normalized fields and therefore to the same naive mass hierarchies as in the previous scenario.

%
%%
%%%
%%%%
%%%%%%%%%%%%%%%%%%%%%%%%%%%%%%%%%%%%%%%%%%%%%%%
%%%%%%%%%%%%%%%%%%%%%%%%%%%%%%%%%%%%%%%%%%%%%%%
\subsection{Backreaction of closed-string moduli}
\label{sec:back}

As explained above, the scalar potential \eqref{eq:V1} is obtained via a two-step approach \cite{Escobar:2015ckf}. First one assumes that all closed-string moduli except $T$ are stabilized to a certain value by a suitable superpotential $W_\text{mod}$ via the condition (\ref{eq:Fvanish}). Second, the full F-term scalar potential is expanded around the inflationary trajectory to derive the leading-order potential in $\Phi$ and $T$. While this procedure gives the correct result for the potential along the inflationary trajectory where the stabilizer is fixed at the origin, it misses important mass terms for the stabilizer field which arise during inflation. In the following we implement the approach of Section \ref{sec:gen} to integrate out the closed-string moduli at tree level to obtain the correct effective potential. As in the toy examples studied earlier, the interaction between moduli and inflaton during inflation leads to tachyonic modes for the stabilizer field which eventually cause the model to fail. This unpleasant effect is ultimately due to the shift symmetries present in the K\"ahler potentials \eqref{eq:D6KK} and \eqref{eq:D6KQ}, as already suggested by the toy models of Section \ref{sec:gen}.

%
%%
%%%
%%%%
%%%%%%%%%%%%%%%%%%%%%%%%%%%%%%%%%%%%%%%%%%%%%%%
\subsubsection*{Backreaction in the Wilson line scenario}
\label{sec:backWilson}

In order to show the importance of the moduli backreaction in the above models of D6-brane inflation let us focus on the scenario in which the Wilson line $\phi = \text{Im}\,\Phi$ is the inflaton candidate. To illustrate the computation of the effective potential it suffices to consider the case of a single complex structure/dilaton modulus $U$ and two K\"ahler moduli $T_\text{v}$ and $T$, where $T_\text v$ parameterizes the complexified overall volume and $T$ is defined by \eqref{eq:D6W}. Taking into account the general expressions \eqref{eq:D6KK} and \eqref{eq:D6KQ} we are lead to the following toy model 
\begin{subequations}\label{eq:toy1}
\begin{align}
K_\text{K} &= - \log \left[\frac16(T_\text{v} + \bar T_\text{v})^3 - \frac12(T_\text{v}+\bar T_\text{v})(T + \bar T)^2 \right] \,, \label{eq:toy11} \\
K_\text{Q} &= -4 \log\left\{ \frac{1}{4} \left[ U + \bar U - \frac{\tilde{Q}}{8} (\Phi + \bar \Phi)^2 \right]\right\}\,, \\
W &= T \Phi + W_\text{mod} (U,T_\text{v})\,,
\end{align}
\end{subequations}
in which we have taken simple choices for the triple intersection numbers and defined $\tilde Q = 2Q/(T_\text{v} + \bar T_\text{v})$. In this parameterization the vacuum of the theory is 
\begin{align}\label{eq:vac1}
\langle \Phi \rangle = \langle T \rangle = 0\,, \quad \langle T_\text{v} \rangle = \mathcal V^{1/3}\,, \quad \langle U \rangle = \mathcal V^{1/2}\,,
\end{align}
where $\mathcal V$ denotes the volume of the compact manifold. The full scalar potential defined by \eqref{eq:toy1} is a complicated expression which is not particularly illuminating. The important parts are however the inflaton couplings at linear and higher order in $U$ and $T_\text{v}$, respectively. Such couplings displace the fields $U$ and $T_\text{v}$ from the vacuum \eqref{eq:vac1} and cause a backreaction into the inflationary system. To see its effect we can expand the scalar potential in terms of this displacement by writing $U = \mathcal V^{1/2} + \delta U(\Phi, T)$ and ${T_\text{v} = \mathcal V^{1/3} + \delta T_\text{v} (\Phi, T)}$, where $\mathcal V$ is treated as a constant fixed by the details of $W_\text{mod}$. Expanding the action and minimizing the result with respect to the fluctuations $\delta U$ and $\delta T_\text{v}$ leads to the following effective potential
\begin{align}\label{eq:V3}\nonumber
V &= \frac12 m^2 \varphi^2 + \left( \frac12 m^2 + m^2 \varphi^2\right) \sigma^2 + \left( \frac12 m^2 -\frac34 m^2 \varphi^2\right) t_1^2 + \left( \frac12 m^2 - \frac34 m^2 \varphi^2\right) t_2^2 \\ 
& \;\;\; + \mathcal O \left(\frac{m \varphi}{\partial_U^2 W_\text{mod}}, \frac{m \varphi}{\partial_U \partial_{T_\text{v}} W_\text{mod}}, \frac{m \varphi}{\partial_{T_\text{v}}^2 W_\text{mod}}  \right)\,,
\end{align}
at quadratic order in the canonically normalized variables. In this derivation we have again used that $W_\text{mod}$ and its first derivatives are small or vanishing in the vacuum, so that the second derivatives define the mass matrix of the closed-string sector. In this case the mass parameter is $m \sim Q^{-1/2} \mathcal V^{-3/4}$. As explained in more detail in \cite{Escobar:2015ckf}, $Q$ scales with the warp factor. Therefore, the mass of the inflaton field can be strongly suppressed compared to all other relevant scales in the theory. Cf.~the more detailed discussion of mass hierarchies in Section \ref{sec:mhier}.

Notice the important difference with respect to the naive result \eqref{eq:V2}: here both components of the stabilizer field are tachyonic during inflation, destabilizing the would-be inflationary trajectory.\footnote{More generally, the axion component of $T$ is always a tachyonic direction during inflation, whereas the saxionic component may or may not be---depending on the specific form of $K$.} This is because the ``remnant" mass terms for the stabilizer found in the two-step procedure of \cite{Escobar:2015ckf}, are actually not present. In particular we find that the last term on the right-hand side of \eqref{eq:V1} is absent, something which is only visible after considering the backreaction of $U$ and $T_\text{v}$ as discussed above.

This back-reacted effective potential can be directly obtained by applying the shortcut discussed in Section \ref{sec:gen}. In particular, the leading-order potential \eqref{eq:V3}, after treating $U$ and $T_\text{v}$ as constants from the beginning,
\begin{align}
K &= - \log \left[\frac43 \mathcal V - \mathcal V^{1/3} (T + \bar T)^2 \right] - 4 \log\left\{ \frac{1}{2} \left[ \mathcal V^{1/2} - \frac{Q}{16 \mathcal V^{1/3}} (\Phi + \bar \Phi)^2 \right]\right\}\,, \\
W &= T \Phi \,.
\end{align}
is identical to the first line of \eqref{eq:V3}. This way, if one is not interested in the corrections suppressed by powers of $m_U$ and $m_{T_\text{v}}$ one can save a lot of effort in computing the back-reacted effective potential. Notice that, from this viewpoint, it is obvious that the moduli dependence of $\tilde Q$ does not play an important role in computing the effective potential. Finally, in this form it is obvious that the cancellation which removes the dangerous negative terms does not take place as expected. What we are left with after backreaction is a variation of the original inflationary theory of \cite{Kawasaki:2000yn}, but with a shift-symmetric K\"ahler potential for the stabilizer field. Actually one can show that in all theories with $K = K(\Phi + \bar \Phi, T + \bar T)$ and the given superpotential the desired mass hierarchy between the inflaton and the stabilizer field cannot be obtained. This applies, in particular, also to the D6-brane inflation scenario in which the inflaton candidate is the B-field, and which fails for the same reason as the case studied above. 

A few comments are in order with respect to these findings. First, via a standard supergravity calculation one can easily verify that including different powers of ${(T+\bar T)}$ in \eqref{eq:toy11} does not solve the problem of the tachyonic directions. Second, the corrections to the leading-order potential in the second line of \eqref{eq:V3}  can never lift the problematic directions. For the theory to be consistent it must be that $m_U, m_{T_\text{v}} \gg H \sim m \varphi$, so that these corrections are always sub-leading. Third, the previous statement is true even in the case when the conditions \eqref{eq:Fvanish} are violated, i.e., if the closed-string moduli are permitted to break supersymmetry. This is more tedious to prove because, in this case, there is no complete decoupling of the heavy fields and the computation of the back-reacted potential is more involved. This analysis has been done in \cite{Buchmuller:2015oma} for a variation of the model at hand, and in \cite{Dudas:2015lga} more generally. There are indeed ``remnant" terms after integrating out $U$ and $T_\text{v}$ in this case, which are proportional to $W_\text{mod}$ and its first derivatives. However, none of them break the shift symmetry of $T$, so the tachyonic directions cannot be lifted.

We conclude that both Wilson line and K\"ahler moduli are unsuitable candidates for stabilizer fields in large field inflationary models, due to the shift symmetry that they display in the K\"ahler potential. Note, however, that such shift symmetries are not fundamental, but an artefact of considering type IIA compactifications with large volumes compared to the string scale. Had we considered compactifications of stringy size, the shift symmetries for the K\"ahler moduli would be generically broken by world-sheet instanton effects and they could in principle serve as stabilizer fields. Nevertheless, the difficulty in that scheme would be to formulate a mechanism that stabilizes the remaining moduli. Indeed, in the large volume limit the source lifting closed-string moduli is a combination of NS and RR fluxes, and implementing the presence of the latter at small volumes remains a challenge. These difficulties are, however, absent if we consider the mirror setup of type IIB compactifications at large volume and small complex structure, as we do in the next section.

%
%%
%%%
%%%%
%%%%%%%%%%%%%%%%%%%%%%%%%%%%%%%%%%%%%%%%%%%%%%%
%%%%%%%%%%%%%%%%%%%%%%%%%%%%%%%%%%%%%%%%%%%%%%%
%%%%%%%%%%%%%%				    %%%%%%%%%%%%%%%%%%%%%%
%%%%%%%%%%%%%%         Section 4           %%%%%%%%%%%%%%%%%%%%%%
%%%%%%%%%%%%%%				    %%%%%%%%%%%%%%%%%%%%%%
%%%%%%%%%%%%%%%%%%%%%%%%%%%%%%%%%%%%%%%%%%%%%%%
%%%%%%%%%%%%%%%%%%%%%%%%%%%%%%%%%%%%%%%%%%%%%%%

\section{Broken shift symmetries and mass hierarchies}
\label{sec:new}

In the previous two sections we have learned that a shift symmetry of the stabilizer field is detrimental to realizing inflation. Whenever the stabilizer field is a K\"ahler modulus in type IIA theories this shift symmetry is inherent to the large volume regime---the desired regime to use ten-dimensional supergravity to treat compactifications with RR fluxes. The mirror dual statement holds for complex structure moduli in type IIB compactifications with O3/O7-planes: shift symmetries are present whenever we consider the large complex structure limit.  However, in such a theory one can explore arbitrary regions of the complex structure moduli space---where the shift symmetries are absent---without sacrificing the ten-dimensional supergravity picture. One may then conceive a model of large-field inflation in which the role of the stabilizer field is played by a complex structure modulus with no shift symmetries, such that the stability problems discussed in the previous section no longer arise. As we discuss below, these fields can have superpotential bilinear couplings to D7-brane Wilson lines, which would then contain the inflaton candidate.

However, even when this obstacle can be overcome in type IIB setups, a bigger one remains: since the warping close to the locus of the brane does not enter the kinetic term of the D7-brane Wilson line in the way that it does for the D6-brane, the necessary mass hierarchies to justify a four-dimensional effective description of single-field inflation cannot be obtained. As explained in more detail below, the mass of the Wilson line axion is generically close to the Kaluza-Klein scale. This seems to render any attempt of realizing chaotic inflation with stabilizer fields in this way futile.

In Section \ref{sec:break} we discuss which effects contribute to the breaking of the shift symmetry in the complex structure sector, at the level of the four-dimensional effective theory. In Section \ref{sec:mhier} we describe how naive attempts to realize inflation in such type IIB setups ultimately fail due to a lack of relative suppression of the inflationary energy scale.

%
%%
%%%
%%%%
%%%%%
%%%%%%%%%%%%%%%%%%%%%%%%%%%%%%%%%%%%%%%%%%%%%%%%
%%%%%%%%%%%%%%%%%%%%%%%%%%%%%%%%%%%%%%%%%%%%%%%%
\subsection{Breaking the shift symmetry in type IIB}
\label{sec:break}

When departing from the large complex structure limit, the mirror of the K\"ahler potential \eqref{eq:D6KK} picks up additional terms which break its shift symmetries. These additional terms can then lift the tachyonic directions encountered in the large volume and large complex structure limits. There are two important sources for this breaking.

%%%%%%%%%%%%%%%%%%%%%%%%%%%%%%%%%%%%%%%%%%%%%%%
\subsubsection*{Closed-string K\"ahler potential}

Consider the closed-string K\"ahler potential for the complex structure moduli, 
\be
K_{\text{cs}}' = - \log \left(i \Pi^{T}\Sigma \Pi \right) = -\log \left[i\left(X^{b}\mathcal{\bar G}_{b}(X) - \bar{X}^{b} \mathcal{G}_{b}(X)\right)\right] \,,
\label{genkah} 
\ee
where $X^b$ and $\mathcal G_b$, $b = 0, \dots, h^{2,1}$, are the entries of the period vector $\Pi$ associated with the holomorphic three-form of the compact manifold. One may then define the four-dimensional fields corresponding to the complex structure deformations by ${z^a = X^a/X^0}$, $a = 1, \dots, h^{2,1}$, and perform the K\"ahler transformation $K_{\text{cs}}' \rightarrow K_{\text{cs}} = K_{\text{cs}}' +{\rm log} |X^0|^2$ to obtain the expression
\be
K_{\text{cs}}  = -\log \left[i\left[(z^a - \bar{z}^{a}) \left(\mathcal{G}_{a} + \mathcal{\bar G}_{a} \right) - 2(\mathcal{G} - \mathcal{\bar G}) \right]\right]\,,
\ee
where ${\mathcal G}(z)$ is understood as a prepotential of a related ${\mathcal N} =2$ theory and \linebreak ${\mathcal{G}_{a}(z) = \partial \mathcal{G}(z)/\partial z^{a}}$.
Expanded around the large complex structure point $z^a \gg 1$ this prepotential can be written as 
\be
\mathcal{G}(z) = \frac{1}{3!}\kappa_{abc} z^{a}z^{b}z^{c} + \frac{1}{2}S_{ab}z^{a}z^{b} + P_{a}z^{a} + Q + \mathcal{G}_\text{exp}\,.
\label{largcxprep}
\ee
Here $\kappa_{abc}$, $S_{ab}$, $P_{a}$ and $Q$ are constants and $\mathcal{G}_\text{exp}$ contains exponentially suppressed contributions which, in the mirror manifold, are identified with world-sheet instantons in the large volume limit. 
This leads to the well-known expression for the K\"ahler potential for the complex structure moduli,
\be
K_{\text{cs}} = - \log \left[\frac{1}{6}\kappa_{abc} \left(U^{a} + \bar{U}^{a}\right)\left(U^{b} + \bar{U}^{b}\right)\big(U^{c} + \bar{U}^{c}\big) + f_{\text{exp}}\right]\,.
\label{largekahz} 
\ee
where we have defined $U^a = i z^a$ in order to connect with the conventions of Section \ref{sec:gen}. In this form the shift symmetry of the imaginary part of the $U^a$ is obvious, broken only by exponentially suppressed contributions.

However, in a regime where $z^a \ll 1$ the exponential corrections in $\mathcal G$ become large and \eqref{largekahz} is no longer a valid expression. Instead, for generic points away from the large complex structure limit, $\mathcal G(z^a)$ is a complicated function which is generally unknown.\footnote{On certain types of manifolds one can expand $\mathcal G$ around other special points in moduli space, like the Landau-Ginzburg point as discussed in \cite{Candelas:1990rm,Berglund:1993ax,Candelas:1993dm} or the conifold point as, for example, discussed in \cite{Huang:2006hq,Garcia-Etxebarria:2014wla,Bizet:2016paj,Blumenhagen:2016bfp}.} Expanding $\mathcal G$ around a point at small complex structure for one modulus $S \in {U^a}$ then yields a more complicated K\"ahler potential. Schematically, one has
\begin{align}\label{eq:Kdestab}
K_\text{cs}(S) &= - \log \Big[\alpha_0 + \alpha_1 (S + \bar S) + \alpha_2 |S|^2 + \alpha_3 (S^2 + \bar S^2) \nonumber \\ 
		& \hspace{1.5cm}+ \alpha_4 (S \bar S^2 + S^2 \bar S) + \alpha_5 (S^3 + \bar S^3) + \dots\Big]\,,
\end{align}
where the coefficients $\alpha_i$ depend on the precise form of the prepotential and its derivatives, as well as the values of the other complex structure moduli. In an effective theory where all $U^a$ except $S$ are stabilized by fluxes, one may treat these as constant parameters. Clearly, some of the terms in \eqref{eq:Kdestab} break the shift symmetry of $S$. Others, like the ones proportional to $\alpha_1$ and $\alpha_4$ can be shown to act destabilizing on the scalar potential for $S$. Therefore, this effect alone may not be sufficient to lift the tachyonic directions associated with the stabilizer field.

%%%%%%%%%%%%%%%%%%%%%%%%%%%%%%%%%%%%%%%%%%%%%%%
\subsubsection*{Open-string K\"ahler potential and superpotential}

One may analogously describe the K\"ahler potential for the K\"ahler deformations of the compactification as follows \cite{Grimm:2004uq}. One first expands the fundamental two-form $J$ and the RR four-form potential $C_4$ as
\begin{align}
J &= v^\alpha\, \omega_\alpha \,, \qquad \omega_\alpha \in H^{2}_+ (\mathcal M_6,\mathbb Z)\,,\nonumber\\
C_4 &= C_\alpha\, \tilde \omega^\alpha\,, \qquad  \tilde \omega^\alpha \in H^{4}_+ (\mathcal M_6,\mathbb Z)\,,
\end{align}
and then describes the four-dimensional chiral coordinates as
\be
T'_\alpha = \frac{1}{2} \mathcal K_{\alpha \beta \gamma} v^\beta v^\gamma - i C_\alpha\,,
\ee
at least in the absence of D3-branes and D7-brane Wilson lines. 
Here
\be
\mathcal K_{\alpha \beta \gamma} = \frac{1}{l^6_\text{s}} \int_{\mathcal M_6} \omega_\alpha \wedge \omega_\beta \wedge \omega_\gamma 
\ee
are the triple intersection numbers of $\mathcal M_6$ and $l_\text{s} = 2\pi \sqrt{\alpha'}$ is the string length. The K\"ahler potential for these moduli is an implicit function of the chiral multiplets,
\be\label{eq:KK}
K_\text{K} = - 2 \log \left[\frac{1}{3!} \mathcal K_{\alpha \beta \gamma} v^\alpha v^\beta v^\gamma\right]   = - \log (\nu(t'_{\alpha}))\,,
\ee
where $\nu(t'_{\alpha})={\mathcal V}^2$ is a homogeneous function of degree three in $t'_{\alpha} = \re T'_\alpha$. 

In the presence of open strings the definition of these chiral coordinates is modified \cite{Jockers:2004yj}. In particular, following \cite{Carta:2016ynn}, one can show that after including D7-brane complex Wilson lines $i\Phi^A = \theta_\beta^A + if^A(U_a) \theta_\alpha^A$ the new chiral variables read
\be\label{eq:That}
T_\alpha = T'_\alpha + \frac{1}{4} \sum_A \frac{\mathcal C^{A}_{\alpha}}{\text{Re}\,  f^{A}(U_a)} \,\Phi^{A} \,\text{Re}\,  \Phi^{A}\,,
\ee
where $A$ runs over the different four-cycles ${\cal S}_A$ wrapped by the D7-branes with Wilson lines, and  $\mathcal C^{A}_{\alpha}$ is a  moduli-independent coupling. More precisely $\mathcal C^{A}_{\alpha} = l_\text{s}^{-4} \int_{\mathcal S_A} \omega_\alpha \wedge \tilde \alpha \wedge \tilde \beta$ is an integer defined in terms of the harmonic one-forms $\tilde \alpha$, $\tilde \beta$ of ${\mathcal S_A}$, see \cite{Carta:2016ynn} for further details. The Wilson lines enter the K\"ahler potential by performing the replacement \eqref{eq:That} in (\ref{eq:KK}). One obtains
\bea\nonumber
K_\text{K}& = &- \log{\left[\nu(t_\alpha)  - \frac{\partial_{t_\alpha} \nu}{16} \sum_A \frac{\mathcal C^{A}_{\alpha}}{\re  f^{A}(U_a)} \left(\Phi^A + \overline{\Phi}^{A}\right)^2 + \dots \right]} \\
& \simeq & -\log{\left[\nu(t_\alpha)  - \frac{{\mathcal V}}{16} \sum_A \frac{\mathcal C^{A}}{\re  f^{A}(U_a)} \left(\Phi^A + \overline{\Phi}^{A}\right)^2 + \dots \right]}\,,
\label{expKK}
\eea
where ${\mathcal C^{A}} = v^\alpha {\mathcal C^{A}_{\alpha}} = l_\text{s}^{-4} \int_{\mathcal S_A} J \wedge \tilde \alpha \wedge \tilde \beta \propto \text{Vol}_{{\mathcal S}_A}^{1/2}$. Most important for our discussion of the shift symmetries is the appearance of complex structure dependent functions $f^A(U^a)$. We can determine these functions whenever Wilson lines appear in the open-string superpotential \cite{Marchesano:2014iea,Carta:2016ynn},
\be
l_\text s W_{\rm D7} =  -\sum_a \frac{1}{\pi l_\text s^2} \int_{\mathcal S_A}  \Omega \wedge  A  = -i \sum_a \theta_\beta^A \left[c_{A\, a} U^a-h_{A}^{a} \mathcal{G}_a \right]  + \theta_{\alpha}^A \left[d_{A}^a\, \mathcal{G}_a - p_{A\, a}\, U^a \right]\,,
\label{supoD7}
\ee
where $(c_{A\, a}, h_{A}^{b}, d_{A}^a,p_{a\, A})$ are moduli-independent integers defined in \cite{Carta:2016ynn}. By imposing that $W$ is holomorphic in the $\Phi^A$ and linear in the $U^a$ we obtain that
\be\label{eq:f}
i f^{A} (U^a) = \frac{ d_{A}^b\mathcal{G}_b - p_{A\, a} U^a\, }{c_{A\, a} U^a}\,.
\ee
In the limit of large complex structure $f^A$ is approximately a linear function of the $U^a$ so that $K_\text K$ respects the shift symmetry of the complex structure sector. Away from that limit, however, higher powers of $U^a$ appear in $f$. In particular, again expanding the prepotential in terms of a single modulus $S \in {U^a}$ leads to the schematic form
\begin{align}
f(S) = a_0 + a_1 S + a_2 S^2 + \dots\,,
\end{align}
where the $a_i$ may once again be regarded as constants once the remaining moduli are stabilized. The appearance of quadratic and higher-order terms in $S$ breaks the shift symmetry. One can check that the resulting mass terms for both components of the stabilizer field can be large enough to lift the tachyonic directions.

%
%%
%%%
%%%%
%%%%%
%%%%%%%%%%%%%%%%%%%%%%%%%%%%%%%%%%%%%%%%%%%%%%%%
%%%%%%%%%%%%%%%%%%%%%%%%%%%%%%%%%%%%%%%%%%%%%%%%
\subsection{Mass hierarchies and challenges for large-field inflation}
\label{sec:mhier}

While the problems involving tachyonic directions in the type IIA scenario seem to be avoidable in the type IIB picture, a new problem arises in this setup. Whenever one describes models of single-field inflation as effective theories of string compactifications, there should be a mass hierarchy of the form
\begin{align}\label{eq:hier}
M_\text{string} > M_\text{KK} > M_\text{cs}, \; M_\text{Kahler} > H_\text{inf}^\star\,,
\end{align}
to guarantee control of the various effective field theories. $H_\text{inf}^\star$ denotes the value of the inflationary Hubble parameter at the point of horizon crossing, i.e., evaluated at the field value $\varphi_\star$ at which the CMB observables are generated. In the large volume regime of a compact manifold with volume $\mathcal V$ it is, therefore, instructive to consider the volume scaling of the different mass scales. For sufficiently isotropic internal manifolds with appropriate fluxes one has, in natural units, $M_\text{string} \propto \mathcal V^{-1/2}$, $M_\text{KK} \propto \mathcal V^{-2/3}$, and $M_\text{cs} \propto N \mathcal V^{-1}$, where $N$ is an $\mathcal O(1)$ coefficient related to the relevant flux quanta \cite{Giddings:2001yu}. Moreover, in K\"ahler moduli stabilization schemes where the $T^\alpha$ break supersymmetry, like KKLT \cite{Kachru:2003aw} or the Large Volume scenario \cite{Balasubramanian:2005zx}, one typically has a mass scale $ \propto W_0 \mathcal V^{-1}$ for many moduli, while the mass scale of others may be suppressed compared to that, meaning $M_\text{Kahler} \propto W_0 \mathcal V^{-3/2}$. Here $W_0$ is usually the vacuum expectation value of the Gukov-Vafa-Witten superpotential. By a tuning of fluxes one can achieve $W_0 \ll 1$, so that a  hierarchy $M_\text{cs} > M_\text{Kahler}$ is possible as well. In the schemes that we consider, i.e., the ones where the K\"ahler moduli do not break supersymmetry, $M_\text{Kahler}$ is typically unrelated to $W_0$, but related to other quantities in $W_\text{mod}$ which may be of $\mathcal O(1)$ or smaller, so that the same structure is preserved \cite{Kallosh:2004yh,Wieck:2014xxa}. 

This very successful scheme ensures the first two inequality signs in \eqref{eq:hier}. So how does the inflationary Hubble parameter scale in the discussed models of D6- or D7-brane inflation? In large-field inflation with a quadratic potential one has, up to $\mathcal O(1)$ factors, $H_\text{inf}^\star = m \varphi_\star$. Here $m$ is the mass of the canonically normalized inflaton field $\varphi$, and it is this parameter that must be suppressed compared to the remaining $M_i$ above. For the case of D6-brane inflation it was shown in \cite{Escobar:2015ckf} that for a D6-brane wrapping a maximally large three-cycle of size ${\mathcal V}^{1/2}$,
\begin{align}
m \propto \frac{1}{Q \mathcal V^{3/4}}\,.
\end{align}
Moreover, it was argued in \cite{Escobar:2015ckf} that in strongly warped regions of the compactification the warp factor enters the coefficient $Q$ linearly. This means that strong warping can suppress $m$ and make up for the lack of volume suppression compared to $M_\text{cs}$ and $M_\text{Kahler}$. Therefore, the hierarchy \eqref{eq:hier} can be achieved and the effective field theories of the model are under control. 

In the case of D7-brane inflation in a type IIB dual theory as outlined in Section \ref{sec:break} the picture is different. Warping does not affect the K\"ahler potential of the D7-brane Wilson line \cite{Marchesano:2008rg}. Expanding the open-string K\"ahler potential as in \eqref{expKK} and computing the canonically normalized mass then leads to
\begin{align}
m \propto \frac{1}{{\mathcal V}^{1/2}\, \text{Vol}_{{\mathcal S}_A}^{1/4}}  \sim  \frac{1}{\mathcal V^{2/3}}\,,
\end{align}
where for simplicity we have assumed that $\text{Vol}_{\mathcal S_A} \sim \mathcal V^{2/3}$, which is obviously the case for compactifications with a single K\"ahler modulus. In the  type IIB case there is no additional suppression of this term because all coefficients that enter are intersection numbers of $\mathcal O(1)$. This means that, at least naively, the inflationary Hubble scale in the type IIB dual description is generically of the same order as the Kaluza-Klein scale and larger than the moduli scales.\footnote{Similar control issues have been encountered in setups involving only closed-string fields, cf.~\cite{Hebecker:2014kva,Blumenhagen:2015qda,Blumenhagen:2015kja}.} This makes a controlled four-dimensional description of single-field inflation impossible.

%
%%
%%%
%%%%
%%%%%%%%%%%%%%%%%%%%%%%%%%%%%%%%%%%%%%%%%%%%%%%
%%%%%%%%%%%%%%%%%%%%%%%%%%%%%%%%%%%%%%%%%%%%%%%
%%%%%%%%%%%%%%				   %%%%%%%%%%%%%%%%%%%%%%
%%%%%%%%%%%%%%         Section 5          %%%%%%%%%%%%%%%%%%%%%%
%%%%%%%%%%%%%%				   %%%%%%%%%%%%%%%%%%%%%%
%%%%%%%%%%%%%%%%%%%%%%%%%%%%%%%%%%%%%%%%%%%%%%%
%%%%%%%%%%%%%%%%%%%%%%%%%%%%%%%%%%%%%%%%%%%%%%%

\section{Summary and discussion}
\label{sec:syc}

We have analyzed compactifications of type II string theories with D-branes with regard to possible realizations of chaotic inflation with a stabilizer field. One can expect such inflationary theories to arise in the four-dimensional effective action because of a bilinear superpotential coupling D-brane Wilson lines and closed-string moduli. We have shown that the type IIA compactification with D6-branes of \cite{Escobar:2015fda,Escobar:2015ckf} admits no stable inflationary trajectories. This becomes evident once the backreaction of heavy closed-string moduli is taken into account. Moreover, we have stressed that integrating out such heavy moduli supersymmetrically is, to leading order, equivalent to treating the moduli as constants in the K\"ahler and superpotential. This provides a simple way to take the leading-order backreaction into account. 

The tachyonic directions in \cite{Escobar:2015fda,Escobar:2015ckf} ultimately arise due to the symmetries of the K\"ahler potential. Specifically, the shift symmetry of the stabilizer field in the large volume limit forbids the necessary large mass terms which stabilize the inflationary trajectory. This observation led us to consider a dual type IIB compactification with D7-branes. While the superpotential coupling the D7-brane Wilson line and closed-string moduli is again bilinear, the K\"ahler potential may be different: Since the stabilizer field is part of the complex structure sector, the pernicious shift symmetry may be broken by considering a point in moduli space away from the large complex structure limit. We have discussed two explicit sources of shift symmetry breaking at such a point which, when combined, can stabilize the potential of the stabilizer field and lift the tachyonic directions.

However, we have also shown that a new problem arises in the type IIB picture which makes our naive attempts at successful inflation fail. The mass of the canonically normalized inflaton field is generically of the same order as the Kaluza-Klein scale, and larger than the scales of moduli stabilization. This is due to the volume scaling of the inflaton mass. All coefficients are of $\mathcal O(1)$ and, in type IIB as opposed to type IIA, strong warping does not suppress the relevant mass scale. Therefore, it is hard to conceive how control over the four-dimensional effective theory could be maintained during inflation. 

We believe that, for this reason, our analysis provides several points worth investigating in the future. First, can a breaking of the shift symmetry of the stabilizer field be achieved in the type IIA picture, where all mass hierarchies are under control? Without sacrificing the large volume regime, possible sources could include $\alpha'$ or $g_\text s$ corrections.
Second, is there a mechanism which could restore the desired mass hierarchies in the type IIB picture, where the tachyonic directions can be lifted? Due to the appearance of the Wilson line modulus in the K\"ahler potential, one may investigate if this is possible in a highly anisotropic region of complex structure moduli space.

%
%%
%%%
%%%%
%%%%%%%%%%%%%%%%%%%%%%%%%%%%%%%%%%%%%%%%%%%%%%%
%%%%%%%%%%%%%%%%%%%%%%%%%%%%%%%%%%%%%%%%%%%%%%%

\subsection*{Acknowledgments}
We would like to thank Michele Cicoli, I\~naki Garc\'ia-Etxebarria, Arthur Hebecker, Liam McAllister, Fabian R\"uhle, Martin Winkler, and Gianluca Zoccarato for stimulating discussions. This work is partially supported by the grants FPA2012-32828 and and FPA2015-65480-P from MINECO, the ERC Advanced Grant SPLE under contract ERC-2012-ADG-20120216-320421 and the grant SEV-2012-0249 of the ``Centro de Excelencia Severo Ochoa" Programme. A.L. is  supported through the FPI grant SVP-2013-067949 and the aid EEBB-I-16-11573 from MINECO. A.L. and F.M. would like to thank UW-Madison for hospitality during completion of this work.

%
%%
%%%
%%%%
%%%%%%%%%%%%%%%%%%%%%%%%%%%%%%%%%%%%%%%%%%%%%%%
%%%%%%%%%%%%%%%%%%%%%%%%%%%%%%%%%%%%%%%%%%%%%%%


\begin{thebibliography}{99}

%\cite{Baumann:2014nda}
\bibitem{Baumann:2014nda} 
  D.~Baumann and L.~McAllister,
  ``Inflation and String Theory,''
  arXiv:1404.2601 [hep-th].
  %%CITATION = ARXIV:1404.2601;%%
  %181 citations counted in INSPIRE as of 06 Jul 2016


%\cite{Westphal:2014ana}
\bibitem{Westphal:2014ana} 
  A.~Westphal,
  ``String cosmology -- Large-field inflation in string theory,''
  Int.\ J.\ Mod.\ Phys.\ A {\bf 30}, no. 09, 1530024 (2015)
  doi:10.1142/S0217751X15300240
  [arXiv:1409.5350 [hep-th]].
  %%CITATION = doi:10.1142/S0217751X15300240;%%
  %22 citations counted in INSPIRE as of 06 Jul 2016


%\cite{Silverstein:2016ggb}
\bibitem{Silverstein:2016ggb} 
  E.~Silverstein,
  ``TASI lectures on cosmological observables and string theory,''
  arXiv:1606.03640 [hep-th].
  %%CITATION = ARXIV:1606.03640;%%


%\cite{Silverstein:2008sg}
\bibitem{Silverstein:2008sg} 
  E.~Silverstein and A.~Westphal,
  ``Monodromy in the CMB: Gravity Waves and String Inflation,''
  Phys.\ Rev.\ D {\bf 78}, 106003 (2008)
  doi:10.1103/PhysRevD.78.106003
  [arXiv:0803.3085 [hep-th]].
  %%CITATION = doi:10.1103/PhysRevD.78.106003;%%
  %440 citations counted in INSPIRE as of 06 Jul 2016


%\cite{McAllister:2008hb}
\bibitem{McAllister:2008hb} 
  L.~McAllister, E.~Silverstein and A.~Westphal,
  ``Gravity Waves and Linear Inflation from Axion Monodromy,''
  Phys.\ Rev.\ D {\bf 82}, 046003 (2010)
  doi:10.1103/PhysRevD.82.046003
  [arXiv:0808.0706 [hep-th]].
  %%CITATION = doi:10.1103/PhysRevD.82.046003;%%
  %404 citations counted in INSPIRE as of 06 Jul 2016


%\cite{Berg:2009tg}
\bibitem{Berg:2009tg} 
  M.~Berg, E.~Pajer and S.~Sjors,
  ``Dante's Inferno,''
  Phys.\ Rev.\ D {\bf 81}, 103535 (2010)
  doi:10.1103/PhysRevD.81.103535
  [arXiv:0912.1341 [hep-th]].
  %%CITATION = doi:10.1103/PhysRevD.81.103535;%%
  %85 citations counted in INSPIRE as of 06 Jul 2016


%\cite{Palti:2014kza}
\bibitem{Palti:2014kza} 
  E.~Palti and T.~Weigand,
  ``Towards large r from [p, q]-inflation,''
  JHEP {\bf 1404}, 155 (2014)
  doi:10.1007/JHEP04(2014)155
  [arXiv:1403.7507 [hep-th]].
  %%CITATION = doi:10.1007/JHEP04(2014)155;%%
  %65 citations counted in INSPIRE as of 06 Jul 2016


%\cite{Retolaza:2015sta}
\bibitem{Retolaza:2015sta} 
  A.~Retolaza, A.~M.~Uranga and A.~Westphal,
  ``Bifid Throats for Axion Monodromy Inflation,''
  JHEP {\bf 1507}, 099 (2015)
  doi:10.1007/JHEP07(2015)099
  [arXiv:1504.02103 [hep-th]].
  %%CITATION = doi:10.1007/JHEP07(2015)099;%%
  %11 citations counted in INSPIRE as of 06 Jul 2016


%\cite{Marchesano:2014mla}
\bibitem{Marchesano:2014mla} 
  F.~Marchesano, G.~Shiu and A.~M.~Uranga,
  ``F-term Axion Monodromy Inflation,''
  JHEP {\bf 1409}, 184 (2014)
  doi:10.1007/JHEP09(2014)184
  [arXiv:1404.3040 [hep-th]].
  %%CITATION = doi:10.1007/JHEP09(2014)184;%%
  %104 citations counted in INSPIRE as of 06 Jul 2016


%\cite{Blumenhagen:2014gta}
\bibitem{Blumenhagen:2014gta} 
  R.~Blumenhagen and E.~Plauschinn,
  ``Towards Universal Axion Inflation and Reheating in String Theory,''
  Phys.\ Lett.\ B {\bf 736}, 482 (2014)
  doi:10.1016/j.physletb.2014.08.007
  [arXiv:1404.3542 [hep-th]].
  %%CITATION = doi:10.1016/j.physletb.2014.08.007;%%
  %69 citations counted in INSPIRE as of 06 Jul 2016


%\cite{Hebecker:2014eua}
\bibitem{Hebecker:2014eua} 
  A.~Hebecker, S.~C.~Kraus and L.~T.~Witkowski,
  ``D7-Brane Chaotic Inflation,''
  Phys.\ Lett.\ B {\bf 737}, 16 (2014)
  doi:10.1016/j.physletb.2014.08.028
  [arXiv:1404.3711 [hep-th]].
  %%CITATION = doi:10.1016/j.physletb.2014.08.028;%%
  %81 citations counted in INSPIRE as of 06 Jul 2016


%\cite{Ibanez:2014kia}
\bibitem{Ibanez:2014kia} 
  L.~E.~Ibanez and I.~Valenzuela,
  ``The inflaton as an MSSM Higgs and open string modulus monodromy inflation,''
  Phys.\ Lett.\ B {\bf 736}, 226 (2014)
  doi:10.1016/j.physletb.2014.07.020
  [arXiv:1404.5235 [hep-th]].
  %%CITATION = doi:10.1016/j.physletb.2014.07.020;%%
  %49 citations counted in INSPIRE as of 06 Jul 2016


%\cite{Hassler:2014mla}
\bibitem{Hassler:2014mla} 
  F.~Hassler, D.~Lust and S.~Massai,
  ``On Inflation and de Sitter in Non-Geometric String Backgrounds,''
  arXiv:1405.2325 [hep-th].
  %%CITATION = ARXIV:1405.2325;%%
  %33 citations counted in INSPIRE as of 06 Jul 2016


%\cite{McAllister:2014mpa}
\bibitem{McAllister:2014mpa} 
  L.~McAllister, E.~Silverstein, A.~Westphal and T.~Wrase,
  ``The Powers of Monodromy,''
  JHEP {\bf 1409}, 123 (2014)
  doi:10.1007/JHEP09(2014)123
  [arXiv:1405.3652 [hep-th]].
  %%CITATION = doi:10.1007/JHEP09(2014)123;%%
  %74 citations counted in INSPIRE as of 06 Jul 2016


%\cite{Franco:2014hsa}
\bibitem{Franco:2014hsa} 
  S.~Franco, D.~Galloni, A.~Retolaza and A.~Uranga,
  ``On axion monodromy inflation in warped throats,''
  JHEP {\bf 1502}, 086 (2015)
  doi:10.1007/JHEP02(2015)086
  [arXiv:1405.7044 [hep-th]].
  %%CITATION = doi:10.1007/JHEP02(2015)086;%%
  %31 citations counted in INSPIRE as of 06 Jul 2016


%\cite{Dudas:2014pva}
\bibitem{Dudas:2014pva} 
  E.~Dudas,
  ``Three-form multiplet and Inflation,''
  JHEP {\bf 1412}, 014 (2014)
  doi:10.1007/JHEP12(2014)014
  [arXiv:1407.5688 [hep-th]].
  %%CITATION = doi:10.1007/JHEP12(2014)014;%%
  %16 citations counted in INSPIRE as of 06 Jul 2016


%\cite{Blumenhagen:2014nba}
\bibitem{Blumenhagen:2014nba} 
  R.~Blumenhagen, D.~Herschmann and E.~Plauschinn,
  ``The Challenge of Realizing F-term Axion Monodromy Inflation in String Theory,''
  JHEP {\bf 1501}, 007 (2015)
  doi:10.1007/JHEP01(2015)007
  [arXiv:1409.7075 [hep-th]].
  %%CITATION = doi:10.1007/JHEP01(2015)007;%%
  %47 citations counted in INSPIRE as of 06 Jul 2016


%\cite{Hayashi:2014aua}
\bibitem{Hayashi:2014aua} 
  H.~Hayashi, R.~Matsuda and T.~Watari,
  ``Issues in Complex Structure Moduli Inflation,''
  arXiv:1410.7522 [hep-th].
  %%CITATION = ARXIV:1410.7522;%%
  %14 citations counted in INSPIRE as of 06 Jul 2016


%\cite{Hebecker:2014kva}
\bibitem{Hebecker:2014kva} 
  A.~Hebecker, P.~Mangat, F.~Rompineve and L.~T.~Witkowski,
  ``Tuning and Backreaction in F-term Axion Monodromy Inflation,''
  Nucl.\ Phys.\ B {\bf 894}, 456 (2015)
  doi:10.1016/j.nuclphysb.2015.03.015
  [arXiv:1411.2032 [hep-th]].
  %%CITATION = doi:10.1016/j.nuclphysb.2015.03.015;%%
  %47 citations counted in INSPIRE as of 06 Jul 2016


%\cite{Ibanez:2014swa}
\bibitem{Ibanez:2014swa} 
  L.~E.~Ibanez, F.~Marchesano and I.~Valenzuela,
  ``Higgs-otic Inflation and String Theory,''
  JHEP {\bf 1501}, 128 (2015)
  doi:10.1007/JHEP01(2015)128
  [arXiv:1411.5380 [hep-th]].
  %%CITATION = doi:10.1007/JHEP01(2015)128;%%
  %36 citations counted in INSPIRE as of 06 Jul 2016


%\cite{Garcia-Etxebarria:2014wla}
\bibitem{Garcia-Etxebarria:2014wla} 
  I.~Garc\'ia-Etxebarria, T.~W.~Grimm and I.~Valenzuela,
  ``Special Points of Inflation in Flux Compactifications,''
  Nucl.\ Phys.\ B {\bf 899}, 414 (2015)
  doi:10.1016/j.nuclphysb.2015.08.008
  [arXiv:1412.5537 [hep-th]].
  %%CITATION = doi:10.1016/j.nuclphysb.2015.08.008;%%
  %23 citations counted in INSPIRE as of 06 Jul 2016


%\cite{Hebecker:2015rya}
\bibitem{Hebecker:2015rya} 
  A.~Hebecker, P.~Mangat, F.~Rompineve and L.~T.~Witkowski,
  ``Winding out of the Swamp: Evading the Weak Gravity Conjecture with F-term Winding Inflation?,''
  Phys.\ Lett.\ B {\bf 748}, 455 (2015)
  doi:10.1016/j.physletb.2015.07.026
  [arXiv:1503.07912 [hep-th]].
  %%CITATION = doi:10.1016/j.physletb.2015.07.026;%%
  %30 citations counted in INSPIRE as of 06 Jul 2016


%\cite{Escobar:2015fda}
\bibitem{Escobar:2015fda} 
  D.~Escobar, A.~Landete, F.~Marchesano and D.~Regalado,
  ``Large field inflation from D-branes,''
  Phys.\ Rev.\ D {\bf 93}, no. 8, 081301 (2016)
  doi:10.1103/PhysRevD.93.081301
  [arXiv:1505.07871 [hep-th]].
  %%CITATION = doi:10.1103/PhysRevD.93.081301;%%
  %14 citations counted in INSPIRE as of 06 Jul 2016


%\cite{Blumenhagen:2015xpa}
\bibitem{Blumenhagen:2015xpa} 
  R.~Blumenhagen, C.~Damian, A.~Font, D.~Herschmann and R.~Sun,
  ``The Flux-Scaling Scenario: De Sitter Uplift and Axion Inflation,''
  Fortsch.\ Phys.\  {\bf 64}, no. 6-7, 536 (2016)
  doi:10.1002/prop.201600030
  [arXiv:1510.01522 [hep-th]].
  %%CITATION = doi:10.1002/prop.201600030;%%
  %10 citations counted in INSPIRE as of 06 Jul 2016


%\cite{Escobar:2015ckf}
\bibitem{Escobar:2015ckf} 
  D.~Escobar, A.~Landete, F.~Marchesano and D.~Regalado,
  ``D6-branes and axion monodromy inflation,''
  JHEP {\bf 1603}, 113 (2016)
  doi:10.1007/JHEP03(2016)113
  [arXiv:1511.08820 [hep-th]].
  %%CITATION = doi:10.1007/JHEP03(2016)113;%%
  %8 citations counted in INSPIRE as of 06 Jul 2016


%\cite{Baume:2016psm}
\bibitem{Baume:2016psm} 
  F.~Baume and E.~Palti,
  ``Backreacted Axion Field Ranges in String Theory,''
  arXiv:1602.06517 [hep-th].
  %%CITATION = ARXIV:1602.06517;%%
  %4 citations counted in INSPIRE as of 06 Jul 2016


%\cite{Kawasaki:2000yn}
\bibitem{Kawasaki:2000yn} 
  M.~Kawasaki, M.~Yamaguchi and T.~Yanagida,
  ``Natural chaotic inflation in supergravity,''
  Phys.\ Rev.\ Lett.\  {\bf 85}, 3572 (2000)
  doi:10.1103/PhysRevLett.85.3572
  [hep-ph/0004243].
  %%CITATION = doi:10.1103/PhysRevLett.85.3572;%%
  %336 citations counted in INSPIRE as of 06 Jul 2016


%\cite{Marchesano:2014iea}
\bibitem{Marchesano:2014iea} 
  F.~Marchesano, D.~Regalado and G.~Zoccarato,
  ``On D-brane moduli stabilisation,''
  JHEP {\bf 1411}, 097 (2014)
  doi:10.1007/JHEP11(2014)097
  [arXiv:1410.0209 [hep-th]].
  %%CITATION = doi:10.1007/JHEP11(2014)097;%%
  %7 citations counted in INSPIRE as of 06 Jul 2016


%\cite{Buchmuller:2014vda}
\bibitem{Buchmuller:2014vda} 
  W.~Buchmuller, C.~Wieck and M.~W.~Winkler,
  ``Supersymmetric Moduli Stabilization and High-Scale Inflation,''
  Phys.\ Lett.\ B {\bf 736}, 237 (2014)
  doi:10.1016/j.physletb.2014.07.024
  [arXiv:1404.2275 [hep-th]].
  %%CITATION = doi:10.1016/j.physletb.2014.07.024;%%
  %21 citations counted in INSPIRE as of 06 Jul 2016


%\cite{Kallosh:2010xz}
\bibitem{Kallosh:2010xz} 
  R.~Kallosh, A.~Linde and T.~Rube,
  ``General inflaton potentials in supergravity,''
  Phys.\ Rev.\ D {\bf 83}, 043507 (2011)
  doi:10.1103/PhysRevD.83.043507
  [arXiv:1011.5945 [hep-th]].
  %%CITATION = doi:10.1103/PhysRevD.83.043507;%%
  %146 citations counted in INSPIRE as of 06 Jul 2016


%\cite{DeWolfe:2005uu}
\bibitem{DeWolfe:2005uu} 
  O.~DeWolfe, A.~Giryavets, S.~Kachru and W.~Taylor,
  ``Type IIA moduli stabilization,''
  JHEP {\bf 0507}, 066 (2005)
  doi:10.1088/1126-6708/2005/07/066
  [hep-th/0505160].
  %%CITATION = doi:10.1088/1126-6708/2005/07/066;%%
  %269 citations counted in INSPIRE as of 06 Jul 2016


%\cite{Camara:2005dc}
\bibitem{Camara:2005dc} 
  P.~G.~Camara, A.~Font and L.~E.~Ibanez,
  ``Fluxes, moduli fixing and MSSM-like vacua in a simple IIA orientifold,''
  JHEP {\bf 0509}, 013 (2005)
  doi:10.1088/1126-6708/2005/09/013
  [hep-th/0506066].
  %%CITATION = doi:10.1088/1126-6708/2005/09/013;%%
  %159 citations counted in INSPIRE as of 06 Jul 2016


%\cite{Palti:2008mg}
\bibitem{Palti:2008mg} 
  E.~Palti, G.~Tasinato and J.~Ward,
  ``WEAKLY-coupled IIA Flux Compactifications,''
  JHEP {\bf 0806}, 084 (2008)
  doi:10.1088/1126-6708/2008/06/084
  [arXiv:0804.1248 [hep-th]].
  %%CITATION = doi:10.1088/1126-6708/2008/06/084;%%
  %33 citations counted in INSPIRE as of 06 Jul 2016


%\cite{Carta:2016ynn}
\bibitem{Carta:2016ynn} 
  F.~Carta, F.~Marchesano, W.~Staessens and G.~Zoccarato,
  ``Open string multi-branched and Kahler potentials,''
  arXiv:1606.00508 [hep-th].
  %%CITATION = ARXIV:1606.00508;%%


%\cite{Grimm:2011dx}
\bibitem{Grimm:2011dx} 
  T.~W.~Grimm and D.~Vieira Lopes,
  ``The N=1 effective actions of D-branes in Type IIA and IIB orientifolds,''
  Nucl.\ Phys.\ B {\bf 855}, 639 (2012)
  doi:10.1016/j.nuclphysb.2011.10.019
  [arXiv:1104.2328 [hep-th]].
  %%CITATION = doi:10.1016/j.nuclphysb.2011.10.019;%%
  %27 citations counted in INSPIRE as of 06 Jul 2016


%\cite{Kerstan:2011dy}
\bibitem{Kerstan:2011dy} 
  M.~Kerstan and T.~Weigand,
  ``The Effective action of D6-branes in N=1 type IIA orientifolds,''
  JHEP {\bf 1106}, 105 (2011)
  doi:10.1007/JHEP06(2011)105
  [arXiv:1104.2329 [hep-th]].
  %%CITATION = doi:10.1007/JHEP06(2011)105;%%
  %26 citations counted in INSPIRE as of 06 Jul 2016


%\cite{Marchesano:2008rg}
\bibitem{Marchesano:2008rg} 
  F.~Marchesano, P.~McGuirk and G.~Shiu,
  ``Open String Wavefunctions in Warped Compactifications,''
  JHEP {\bf 0904}, 095 (2009)
  doi:10.1088/1126-6708/2009/04/095
  [arXiv:0812.2247 [hep-th]].
  %%CITATION = doi:10.1088/1126-6708/2009/04/095;%%
  %49 citations counted in INSPIRE as of 06 Jul 2016


%\cite{Buchmuller:2014pla}
\bibitem{Buchmuller:2014pla} 
  W.~Buchmuller, E.~Dudas, L.~Heurtier and C.~Wieck,
  ``Large-Field Inflation and Supersymmetry Breaking,''
  JHEP {\bf 1409}, 053 (2014)
  doi:10.1007/JHEP09(2014)053
  [arXiv:1407.0253 [hep-th]].
  %%CITATION = doi:10.1007/JHEP09(2014)053;%%
  %26 citations counted in INSPIRE as of 06 Jul 2016


%\cite{Kallosh:2004yh}
\bibitem{Kallosh:2004yh} 
  R.~Kallosh and A.~D.~Linde,
  ``Landscape, the scale of SUSY breaking, and inflation,''
  JHEP {\bf 0412}, 004 (2004)
  doi:10.1088/1126-6708/2004/12/004
  [hep-th/0411011].
  %%CITATION = doi:10.1088/1126-6708/2004/12/004;%%
  %223 citations counted in INSPIRE as of 06 Jul 2016


%\cite{Wieck:2014xxa}
\bibitem{Wieck:2014xxa} 
  C.~Wieck and M.~W.~Winkler,
  ``Inflation with Fayet-Iliopoulos Terms,''
  Phys.\ Rev.\ D {\bf 90}, no. 10, 103507 (2014)
  doi:10.1103/PhysRevD.90.103507
  [arXiv:1408.2826 [hep-th]].
  %%CITATION = doi:10.1103/PhysRevD.90.103507;%%
  %16 citations counted in INSPIRE as of 06 Jul 2016


%\cite{Kachru:2003aw}
\bibitem{Kachru:2003aw} 
  S.~Kachru, R.~Kallosh, A.~D.~Linde and S.~P.~Trivedi,
  ``De Sitter vacua in string theory,''
  Phys.\ Rev.\ D {\bf 68}, 046005 (2003)
  doi:10.1103/PhysRevD.68.046005
  [hep-th/0301240].
  %%CITATION = doi:10.1103/PhysRevD.68.046005;%%
  %2288 citations counted in INSPIRE as of 06 Jul 2016


%\cite{Balasubramanian:2005zx}
\bibitem{Balasubramanian:2005zx} 
  V.~Balasubramanian, P.~Berglund, J.~P.~Conlon and F.~Quevedo,
  ``Systematics of moduli stabilisation in Calabi-Yau flux compactifications,''
  JHEP {\bf 0503}, 007 (2005)
  doi:10.1088/1126-6708/2005/03/007
  [hep-th/0502058].
  %%CITATION = doi:10.1088/1126-6708/2005/03/007;%%
  %564 citations counted in INSPIRE as of 06 Jul 2016


%\cite{Balasubramanian:2004uy}
\bibitem{Balasubramanian:2004uy} 
  V.~Balasubramanian and P.~Berglund,
  ``Stringy corrections to Kahler potentials, SUSY breaking, and the cosmological constant problem,''
  JHEP {\bf 0411}, 085 (2004)
  doi:10.1088/1126-6708/2004/11/085
  [hep-th/0408054].
  %%CITATION = doi:10.1088/1126-6708/2004/11/085;%%
  %158 citations counted in INSPIRE as of 06 Jul 2016


%\cite{Westphal:2006tn}
\bibitem{Westphal:2006tn} 
  A.~Westphal,
  ``de Sitter string vacua from Kahler uplifting,''
  JHEP {\bf 0703}, 102 (2007)
  doi:10.1088/1126-6708/2007/03/102
  [hep-th/0611332].
  %%CITATION = doi:10.1088/1126-6708/2007/03/102;%%
  %69 citations counted in INSPIRE as of 06 Jul 2016


%\cite{Dong:2010in}
\bibitem{Dong:2010in} 
  X.~Dong, B.~Horn, E.~Silverstein and A.~Westphal,
  ``Simple exercises to flatten your potential,''
  Phys.\ Rev.\ D {\bf 84}, 026011 (2011)
  doi:10.1103/PhysRevD.84.026011
  [arXiv:1011.4521 [hep-th]].
  %%CITATION = doi:10.1103/PhysRevD.84.026011;%%
  %90 citations counted in INSPIRE as of 06 Jul 2016


%\cite{Buchmuller:2015oma}
\bibitem{Buchmuller:2015oma} 
  W.~Buchmuller, E.~Dudas, L.~Heurtier, A.~Westphal, C.~Wieck and M.~W.~Winkler,
  ``Challenges for Large-Field Inflation and Moduli Stabilization,''
  JHEP {\bf 1504}, 058 (2015)
  doi:10.1007/JHEP04(2015)058
  [arXiv:1501.05812 [hep-th]].
  %%CITATION = doi:10.1007/JHEP04(2015)058;%%
  %30 citations counted in INSPIRE as of 06 Jul 2016


%\cite{Dudas:2015lga}
\bibitem{Dudas:2015lga} 
  E.~Dudas and C.~Wieck,
  ``Moduli backreaction and supersymmetry breaking in string-inspired inflation models,''
  JHEP {\bf 1510}, 062 (2015)
  doi:10.1007/JHEP10(2015)062
  [arXiv:1506.01253 [hep-th]].
  %%CITATION = doi:10.1007/JHEP10(2015)062;%%
  %12 citations counted in INSPIRE as of 06 Jul 2016


%\cite{Candelas:1990rm}
\bibitem{Candelas:1990rm} 
  P.~Candelas, X.~C.~De La Ossa, P.~S.~Green and L.~Parkes,
  ``A Pair of Calabi-Yau manifolds as an exactly soluble superconformal theory,''
  Nucl.\ Phys.\ B {\bf 359}, 21 (1991).
  doi:10.1016/0550-3213(91)90292-6
  %%CITATION = doi:10.1016/0550-3213(91)90292-6;%%
  %654 citations counted in INSPIRE as of 06 Jul 2016


%\cite{Berglund:1993ax}
\bibitem{Berglund:1993ax} 
  P.~Berglund, P.~Candelas, X.~De La Ossa, A.~Font, T.~Hubsch, D.~Jancic and F.~Quevedo,
  ``Periods for Calabi-Yau and Landau-Ginzburg vacua,''
  Nucl.\ Phys.\ B {\bf 419}, 352 (1994)
  doi:10.1016/0550-3213(94)90047-7
  [hep-th/9308005].
  %%CITATION = doi:10.1016/0550-3213(94)90047-7;%%
  %52 citations counted in INSPIRE as of 06 Jul 2016


%\cite{Candelas:1993dm}
\bibitem{Candelas:1993dm} 
  P.~Candelas, X.~De La Ossa, A.~Font, S.~H.~Katz and D.~R.~Morrison,
  ``Mirror symmetry for two parameter models. 1.,''
  Nucl.\ Phys.\ B {\bf 416}, 481 (1994)
  doi:10.1016/0550-3213(94)90322-0
  [hep-th/9308083].
  %%CITATION = doi:10.1016/0550-3213(94)90322-0;%%
  %210 citations counted in INSPIRE as of 06 Jul 2016


%\cite{Huang:2006hq}
\bibitem{Huang:2006hq} 
  M.~x.~Huang, A.~Klemm and S.~Quackenbush,
  ``Topological string theory on compact Calabi-Yau: Modularity and boundary conditions,''
  Lect.\ Notes Phys.\  {\bf 757}, 45 (2009)
  [hep-th/0612125].
  %%CITATION = HEP-TH/0612125;%%
  %103 citations counted in INSPIRE as of 06 Jul 2016


%\cite{Bizet:2016paj}
\bibitem{Bizet:2016paj} 
  N.~Cabo Bizet, O.~Loaiza-Brito and I.~Zavala,
  ``Mirror quintic vacua: hierarchies and inflation,''
  arXiv:1605.03974 [hep-th].
  %%CITATION = ARXIV:1605.03974;%%
  %2 citations counted in INSPIRE as of 06 Jul 2016


%\cite{Blumenhagen:2016bfp}
\bibitem{Blumenhagen:2016bfp} 
  R.~Blumenhagen, D.~Herschmann and F.~Wolf,
  ``String Moduli Stabilization at the Conifold,''
  arXiv:1605.06299 [hep-th].
  %%CITATION = ARXIV:1605.06299;%%
  %1 citations counted in INSPIRE as of 06 Jul 2016


%\cite{Grimm:2004uq}
\bibitem{Grimm:2004uq} 
  T.~W.~Grimm and J.~Louis,
  ``The Effective action of N = 1 Calabi-Yau orientifolds,''
  Nucl.\ Phys.\ B {\bf 699}, 387 (2004)
  doi:10.1016/j.nuclphysb.2004.08.005
  [hep-th/0403067].
  %%CITATION = doi:10.1016/j.nuclphysb.2004.08.005;%%
  %259 citations counted in INSPIRE as of 06 Jul 2016


%\cite{Jockers:2004yj}
\bibitem{Jockers:2004yj} 
  H.~Jockers and J.~Louis,
  ``The Effective action of D7-branes in N = 1 Calabi-Yau orientifolds,''
  Nucl.\ Phys.\ B {\bf 705}, 167 (2005)
  doi:10.1016/j.nuclphysb.2004.11.009
  [hep-th/0409098].
  %%CITATION = doi:10.1016/j.nuclphysb.2004.11.009;%%
  %168 citations counted in INSPIRE as of 06 Jul 2016


%\cite{Giddings:2001yu}
\bibitem{Giddings:2001yu} 
  S.~B.~Giddings, S.~Kachru and J.~Polchinski,
  ``Hierarchies from fluxes in string compactifications,''
  Phys.\ Rev.\ D {\bf 66}, 106006 (2002)
  doi:10.1103/PhysRevD.66.106006
  [hep-th/0105097].
  %%CITATION = doi:10.1103/PhysRevD.66.106006;%%
  %1562 citations counted in INSPIRE as of 06 Jul 2016


%\cite{Blumenhagen:2015qda}
\bibitem{Blumenhagen:2015qda} 
  R.~Blumenhagen, A.~Font, M.~Fuchs, D.~Herschmann and E.~Plauschinn,
  ``Towards Axionic Starobinsky-like Inflation in String Theory,''
  Phys.\ Lett.\ B {\bf 746}, 217 (2015)
  doi:10.1016/j.physletb.2015.05.001
  [arXiv:1503.01607 [hep-th]].
  %%CITATION = doi:10.1016/j.physletb.2015.05.001;%%
  %21 citations counted in INSPIRE as of 06 Jul 2016


%\cite{Blumenhagen:2015kja}
\bibitem{Blumenhagen:2015kja} 
  R.~Blumenhagen, A.~Font, M.~Fuchs, D.~Herschmann, E.~Plauschinn, Y.~Sekiguchi and F.~Wolf,
  ``A Flux-Scaling Scenario for High-Scale Moduli Stabilization in String Theory,''
  Nucl.\ Phys.\ B {\bf 897}, 500 (2015)
  doi:10.1016/j.nuclphysb.2015.06.003
  [arXiv:1503.07634 [hep-th]].
  %%CITATION = doi:10.1016/j.nuclphysb.2015.06.003;%%
  %26 citations counted in INSPIRE as of 06 Jul 2016


\end{thebibliography}
\end{document}